# Single photon detection up to 2 μm in pair of parallel microstrips based on NbRe ultrathin films

C. Cirillo[1,*], M. Ejrnaes[2], P. Ercolano[3], C. Bruscino[3], A. Cassinese[3], D. Salvoni[4], C. Attanasio[5], G.P. Pepe[3], and L. Parlato[3]

[1]*CNR-SPIN, c/o Università di Salerno, Via Giovanni Paolo II, 132, 84084 Fisciano (Sa), Italy*
[2]*CNR-SPIN, Via Campi Flegrei, 34, 80078 Pozzuoli (Na), Italy*
[3]*Dipartimento di Fisica "E. Pancini", Università degli Studi di Napoli Federico II, 80125 Napoli, Italy*
[4]*Photon Technology Italy srl, via Giacinto Gigante 174, 80128 Napoli Italy*
[5]*Dipartimento di Fisica "E. R. Caianiello", Università degli Studi di Salerno, 84084, Fisciano (Sa), Italy*

**Abstract**

Superconducting Microstrip Single Photon Detectors (SMSPDs) are increasingly attracting the interest of the scientific community as a new platform for large area detectors with unprecedented advantaged in terms of fabrication. However, while their operativity at the telecommunication wavelength was achieved, working beyond 1.55 μm is challenging. Here, we experimentally demonstrate single-photon operation of NbRe microstrips at wavelengths of 1.55 and 2 μm. The devices are structured as pairs of parallel microstrips with widths ranging from 1.4 to 2.2 μm and lengths from 5 to 10 μm. This innovative design may assure large sensitive areas, without affecting the kinetic inductance, namely the time performance of the detectors. The results are discussed in the framework of the hot-spot two-temperature model.

(*) Corresponding Author: carla.cirillo@spin.cnr.it

**Introduction**

Twenty years after their introduction [1], Single Photon Detectors based on superconducting materials are still the state of the art technology due to their unprecedented performances [2-4]. Despite being the core of photon-based quantum technology, such as communication, computation and sensing, they represent an active field of research [5]. Many challenges are still open in order to optimize the performance of these devices for specific fields. For example, different applications require large area coverage, such as dark matter research or Light detection and ranging (LIDAR) [6-9]. For these reasons, efforts are made to increase the widths of the superconducting wires responsible of the detection from the nanoscale, as in Superconducting Nanowire Single Photon Detectors (SNSPDs) [3], to micrometric scale, as in the recently proposed Superconducting Microstrips Single Photon Detectors (SMSPDs) [10]. Both devices are based on ultrathin superconducting films typically patterned in forms of meanders, with thickness ($d$) of the order of the superconducting coherence length ($\xi$) operating below the critical temperature ($T_c$) and at bias currents very close to the critical current density ($J_c$). SNSPDs consist of wires of width in the 30-100 nm range. They were the first to be proposed and they currently are the state-of-the-art devices. Several mechanisms were proposed to model the detection process in SNSPD [11]. In the very first picture [12], an absorbed photon locally heats and thus destroys the superconducting state, creating a normal region, called hot-spot (HS). The HS expands in the nanowire with a dynamics that depends on the microscopical material parameters, until the bias current density is forced to flow in a such reduced area that $J_c$ is overcome and the device switches to the normal state. Summarising, a voltage pulse is produced by an absorbed photon. However, this picture is oversimplified, since in a more realistic refined model [13], the HS is replaced by a finite region of weakened superconductivity, in which less Cooper pairs should carry the same current. As a result their velocity can overcome the critical one, and, as a result, the superconducting state can collapse. Both these models account for the dependence of the efficiency on the bias current, since for a given photon energy the number of quasiparticles present in the nanowire increases with increasing $J$. Therefore, operating at larger $J$, makes the system closer to the transition point, and consequently more sensitive. However, to account for other experimental observations, such as the dependence of the efficiency on the bias current, or on the position of the incident photon, the role of superconducting vortices, both single ones [14,15] and vortex-antivortex pairs (VAP) [16] was invoked. In particular, it was suggested that the arrival of a photon may either decrease the energy barrier for vortex entry or may lead to the formation and subsequent breaking of a VAP due to the Lorentz force generated by the bias current. Moreover, different processes can take place at the same time. The study of the dark count rate, namely events not generated by an arriving photon, as a function of Ibias can shed a light on the processed involved [17]. On top of that, as discussed in several works, one should mention that the material parameters, as well as the device design, play a key role in the determination of the dominant detection mechanism [45]. In this sense, a paradigm shift is represented by the work of Vodolazov [10], where it was suggested that micrometric strips can in principle work in the single photon regime as long as the bias current is uniformly distributed and the energy transfer from the photon to the electronic system is efficient and confined in a small region. These requirements set specific constraints to different microscopical parameters such as resistivity, penetration depth, electronic heat capacity, relaxation rates. In the case of SMSPDs, operativity as single photon detector at 1.55 μm [18-21], and only recently, up to 2 μm [22] was demonstrated. This has renewed the interest in the realization of detectors based on high-temperature superconductors, such as YBCO and $MgB_2$ [23-26]. In fact, while these materials can help to increase the operation temperature of the detectors, their superconducting properties typically suffer from the nanopatterning procedure. Moreover, there is an increasing demand of devices capable of detecting low-energy photon [27]. In fact, working beyond 1.55 μm opens the path to a wide range of applications, as in the atmosphere monitoring of pollutants and greenhouse gases with LIDARs [16, 28], in free space communications, or in the space-ground integrated quantum network [29]. However, this frequency range requires the use of low-gap superconductors such as amorphous ones [28, 30-32], but with the drawback of operation temperatures in the mK regime. Alternatively, crystalline superconductors were also employed, such as NbTiN, with extremely reduced thickness and width, demanding high-resolution nanopattering [33], or by changing the phase of NbN [34] or irradiating it [32]. It is therefore clear that a material-oriented research is crucial for obtaining devices with these demanding performances. For all the reasons previously discussed, the realisation of large-area detectors based on nanowires working at wavelengths longer than 1.55 μm and temperatures accessible by standard cryocooler is desirable, but still challenging.

In this work, we show that NbRe-based superconducting detectors [35-37] can meet all the-above-mentioned requirements. At this purpose we realized devices with an innovative microstrip layout which consists of a pair of parallel strips, 1.4 – 2.2 μm wide. This design can be considered as the building block for the realization of more complex structures based on short microstrips arranged in a parallel/series configuration. This solution combines the possibility of having large-area detectors, while using relatively short strips, thus strongly reducing the kinetic inductance, $L_K$, of the device. The superconducting single photon detector (SSPD) is inserted into a bias and readout circuit with a load resistance $R_L$ placed in parallel to the device. The electrical equivalent of the SSPD is an inductor with kinetic inductance $L_k$, in series with a parallel block made of a resistor, $R_n$, the resistance of the normal domain and a switch. When the strip is in the superconducting state, the switch is closed and shunts the resistor; after the arrival of a particle, the switch opens and adds the resistance $R_n$ in series to the kinetic inductance. $L_k$ depends on the geometry of the device, namely width ($w$), length ($l$) and thickness ($d$), and on the material parameter through the London penetration depth: $L_K = \mu_0 \lambda_L^2 \frac{l}{wd}$. Moreover, the rise time, $\tau_{rise}$, and the decay time, $\tau_{decay}$, of the output voltage pulse are both proportional to the kinetic inductance of the strip [38] as $\tau_{rise} = \frac{L_k}{R_N+R_L}$ and $\tau_{decay} = \frac{L_k}{R_L}$. A decrease of $L_K$, produces a reduction of the recovery time and an increase of the maximum count rate of the detector. The dark count rates and mechanisms involved in the detection process in this kind of structures were preliminarily studied in a previous work [37]. Here, we demonstrate that the devices can operate in the single photon regime at λ larger than telecom wavelength, up to 2 μm, and at temperatures of 1.6 K, easily achievable by commercial cryocooler systems. These results represent significant progress for the realisation of large-area detectors working beyond the telecommunication wavelength at a relatively high temperature. The potential of NbRe ultrathin films for the realisation of SMSPDs with improved performances are delineated in the discussion, where also the perspectives of this work are reported.

**Experiment**

$Nb_{0.18}Re_{0.82}$ (hereafter NbRe) ultrathin films with thickness, $d$, of 4 nm have been deposited on $Si/SiO_x$ substrates by DC magnetron sputtering in ultra-high vacuum ($P \sim 10^{-8}$ mbar) at room temperature. The deposition was performed in an Ar pressure of $4 \cdot 10^{-3}$ mbar at a growth rate of 0.3 nm/s [39]. The NbRe surface was then protected by a 2-nm-thick Al cap layer. The samples were patterned through optical lithography by using a microprinter. Devices with different nominal widths ranging from 1.4 μm to 2.2 μm, and lengths from 5 μm to 12.5 μm were fabricated. The devices are pairs of parallel strips, a test geometry that can be replicated in the future to realize series of pairs of parallel strips in order to gradually increase the detection area without affecting the values of the kinetic inductance. All the corners of the devices are rounded to reduce the current crowding effect. All the fabricated devices, their names and geometrical characteristics, as well as their superconducting properties are summarized in Table 1. In Figure 1 a layout and a microscope photograph of the device D1 are shown.

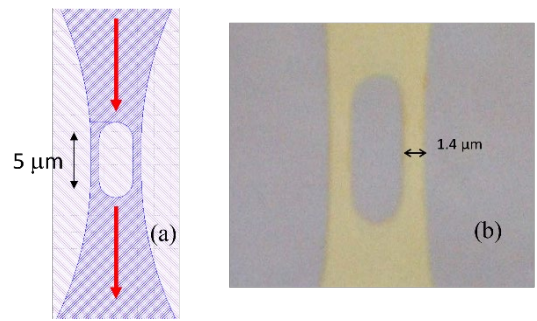

**Figure 1.** Layout (a) and microscope photograph (b) of the device D1, consisting in a single pair of parallel strips. The black arrows indicate the length, $L$, and the width, $w$, of the strips at the narrowest point. The red arrows indicate the direction of the current flow.

Electrical transport measurements were performed in a liquid $^4$He cryostat with the samples mounted on a dipstick equipped with a calibrated thermometer, by using a standard four-wire configuration. The superconducting critical temperature, $T_c$, was resistively measured and defined at the midpoint of the resistive transition. For all devices, we estimated $T_c$ =5.15 K, in agreement with the results reported on unstructured films [40]. A typical current-voltage (I–V) characteristic measured at $T$=4.2 K on the device D1 is shown in Figure 2.

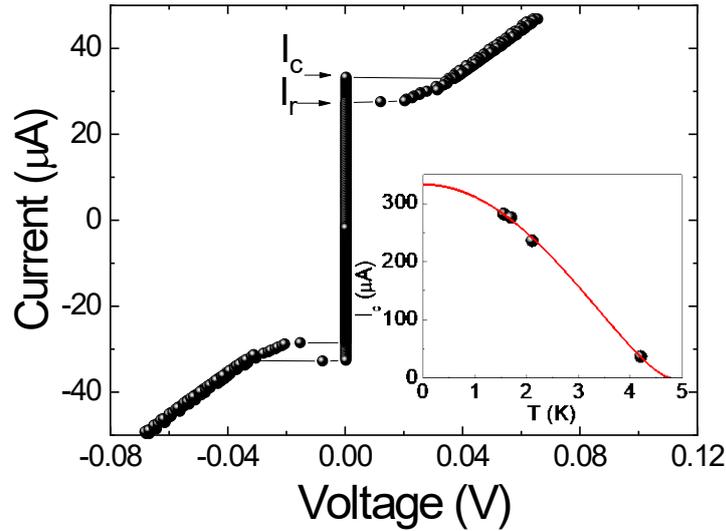

**Figure 2.** I-V curve at T=4.2 K for the device D1. $I_c$ and $I_r$, are the switching and the retrapping current, respectively. Inset: temperature dependence of the critical current, $I_c(T)$. The red line is the fitting curve obtained by using the Bardeen expression.

As the bias current, $I_b$, is swept from zero to higher values, the strip exhibits an abrupt transition from the superconducting (zero voltage) to the normal (finite voltage) state at the switching current, $I_c$. The strip persist in the normal state until the current is reduced below the re-trapping current, $I_r$, when the strip becomes again superconducting. This gives rise to a hysteretic current-voltage characteristic, that is a fingerprint of the presence of a hot-spot in the superconducting strip. Moreover, it assures a clear bistable switching from the superconducting to the dissipative state, which is crucial for the operation of the strip as a SMSPD. In table 1, we report the difference between the switching and the retrapping currents ($I_c$ - $I_r$) as quality parameter.

For all the investigated devices the values the critical current density, $J_c=I_c/wd= I_c/A$ at T=4.2 K are in the range 0.1–0.4 MA/cm$^2$. The behaviour of $I_c$ as a function of the temperature for the device D1 is reported in the inset of Figure 2. The experimental data are fitted according to the Bardeen expression [41] $I_c(T) = I_0 \left(1 - \left(\frac{T}{T_c}\right)^2\right)^{3/2}$, with $I_0 = (360 \pm 10)\ \mu A$ and $T_c = (4.8 \pm 0.1)\ K$. The value of $T_c$ is slightly lower than the measured value, probably due to the fact the previous formula does not reproduce the critical current values at higher temperatures, when vortex and antivortex pairs start to dissociate [37]. Moreover, at the lowest temperature reachable with our experimental set-up, approximatively 1.6 K, $I_c$ is not yet saturated. We will come back to this point and to the related consequences when analysing the detection performances of the devices.

In the following we focus on the device showing the larger critical current values, namely D1 (see Table 1). In fact, it is known that the single photon regime in microstrips can be obtained only if the ratio between the $J_c$ and the depairing current, $J_{dp}$, is beyond a specific threshold which depends on the parameter $\gamma=(8\pi^2/5)\ C_e/C_p$ [10]. $\gamma$ measures the ability of the material to maintain as much as possible the photon energy in a spatially confined region of excitations in the electron system (here $C_e$ and $C_p$ are heat capacities of the electron and photon system, respectively). For our NbRe films at $T=T_c$ it is $\gamma \approx 15$ [42] and we estimated that single photon regime was achieved only if the ratio $J_c/J_{dp}$ was above 0.21 [36].

The devices were illuminated through a multimode fiber that couples light from room temperature to the refrigerated part where it ends 52 mm above the SMSPDs and emits a cone of light to ensure uniform illumination. The stripes are electrically connected through a small printed circuit to a coaxial cable that goes up to room temperature where a bias-tee (Mini-circuits ZFBT-6GWþ) allows to apply a noise filtered DC bias from custom low-noise electronics. It also amplifies the SMSPD signal pulses using an RF amplifier (Mini-circuits ZFL- 1000LN) before registering them on an oscilloscope or counting them. In order to avoid latching, which could affect our microstrips due to their short length, we inserted a 470 nH series inductance and a 333 Ω parallel resistor in the measurement circuit [43]. Two diode lasers at different wavelengths, λ, were used. The light emitted from the first laser has λ=1550 nm and a continuous power of 1 mW. In this case, to obtain a pulsed beam, the pulse/pattern generator mod.81104A by Hewlett-Packard has been used. The laser is fiber coupled and attenuated in two stages, the first one using an electronic variable optical attenuator by Thorlabs and the second stage of attenuation is done using a section of free space beam propagation where we can insert neutral density filters in the beam path. The other laser (FPL2000S by Thorlabs) with λ=2 μm has a control unit which allows the setting of the operation current flowing through it, and then of the output power. However, these values are an overestimation of the power that actually reaches the sample, since losses take place at the FC/PC connectors between the fiber and the insert.

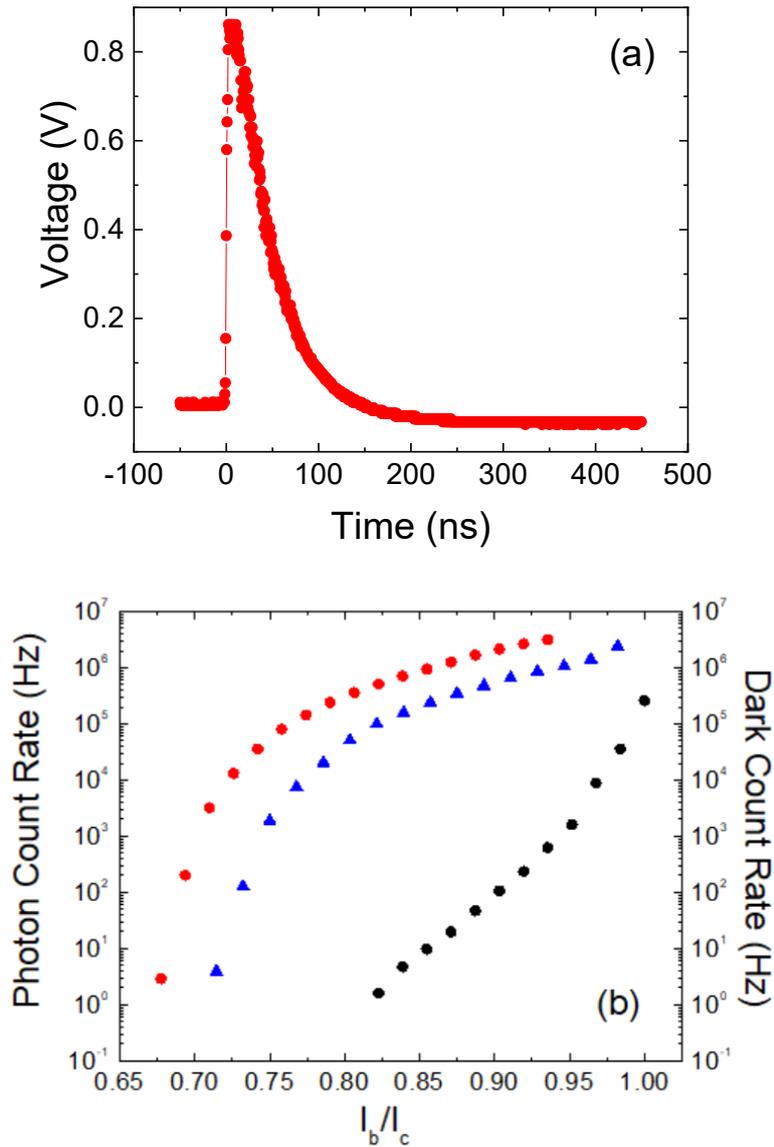

**Figure 3** – (a) Typical photon response pulse for the detector D1 at T= 1.57 K and $I_b$=200 µA at 1.5 µm. (b) Dark count rate (black circles) and photon count rate at 1.55 µm (red circles) and at 2.00 µm (blue triangles) for the device D1 at T=1.57 K.

A typical photon response pulse for the detector D1 at T= 1.57 K, $I_b$=200 µA, and at 1.5 µm is reported in Fig. 3(a). The main goal of this work is to demonstrate the single-photon sensitivity of NbRe microwires beyond the telecommunication wavelength, up to 2 µm. The results in terms of the photon count rate (PCR, left scale) and the dark count rate (DCR, right scale) at T=1.57 K ($T/T_c$=0.31) using laser pulses at 1.55 µm (2 µm) are shown in Fig. 3(b) by red (blue) symbols for the device D1. A nearly saturated PCR is obtained at both wavelengths, probably due to the geometry of the samples, since the region sensitive to light can expand into the tapered edges at higher bias current. Better performances in this sense could in principle be achieved by reducing the operation temperature, since working below 1 K can assure higher $I_c$ values (see inset of Fig. 2), with obvious benefits in terms of PCR and DCR [45]. Similar results can be achieved by further reducing the films thickness, which will also require measuring at lower temperatures. Concerning false events, the dark count rates show an exponentially increase with $I_b$. A change in the DCR slope at $I_b/I_c$=0.94, related to the background and the intrinsic contribution, is clearly visible.

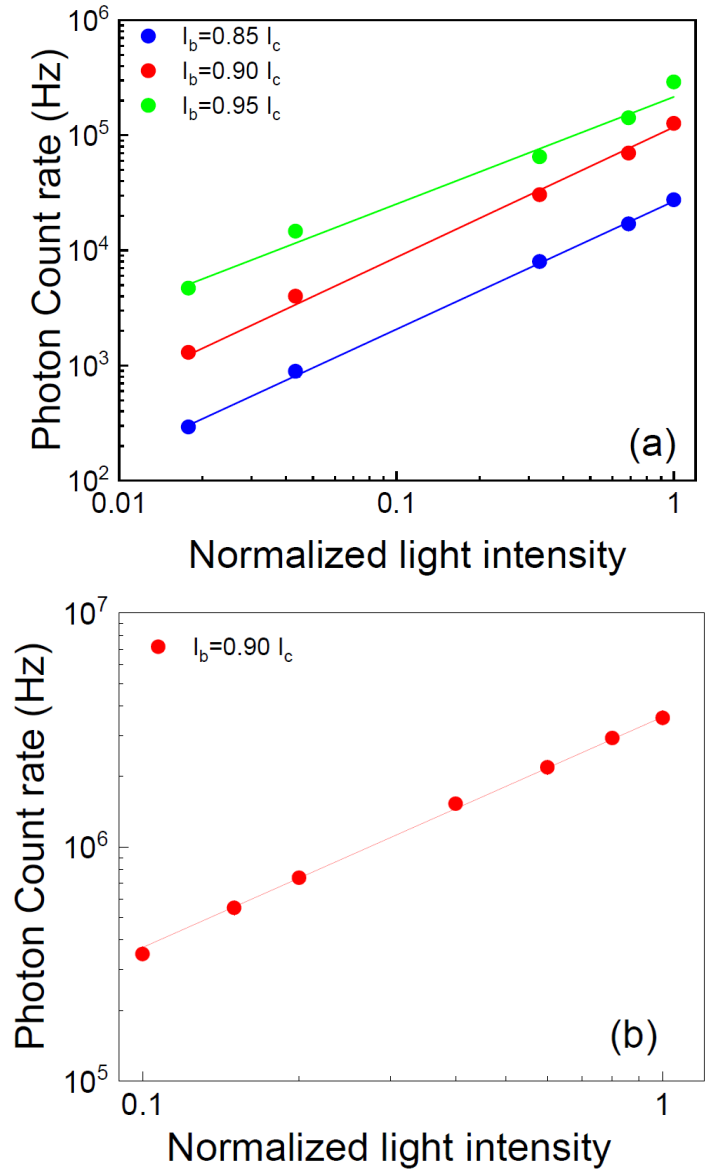

**Figure 4.** Photon count rate as a function of light intensity for the device D1 at T=1.57 K. The lines represent the fitting curves of the experimental data at (a) 1.55 μm as a function of the bias current, and at (b) 2.0 μm for $I_b/I_c$=0.90.

Finally, by measuring the PCR as a function of the light attenuation for both the analysed wavelengths, we verified that the devices were working in the single-photon counting regime. By assuming that the photon statistics has a Poissonian distribution, for a mean number of $m$ photons per pulse, the probability P(n) of detecting n photons from a given pulse is proportional to P(n)≅$e^{-m}(m)^n/n!$ By attenuating the radiation fluence to reduce the total number of incident photons on the detector the condition m « 1 (corresponding to having on average much less than one photon per pulse) is achieved. In this case it is: P(n)≅mn/n! Since the energy of the pulse is proportional to $m$, the photon count rate in the detection regime of n-photons exhibits a power-law dependence on the mean pulse energy where the exponent is $n$. Therefore, if the detection process is of the single-photon type, the photon count rate is linearly proportional to the photon flux [1]. In Fig. 4a, the result obtained as a function of the ratio $I_b/I_c$ is reported at λ=1.55 μm. The experimental data, which spread over two orders of magnitude in a wide range of $I_b$, were successfully fitted by the dependence $PCR = a \cdot P^b$, where $a$ and $b$ are free fitting parameters (see lines in Fig. 4a). The same analysis was performed at λ=2.0 μm by fixing $I_b/I_c$=0.90, as shown in Fig. 4b. Also in this case the data are satisfactory reproduced by a linear fit. As reported in Table 2, the value of $b$ is very close to unity for all the analysed currents and at both wavelengths, demonstrating that the detection probability depends linearly on the average number of incident photons. Therefore, we can conclude that the device works in the single photon regime at λ=1.55 μm as well as at λ=2.0 μm.

**Discussion**

Before commenting on the results obtained for device D1 under illumination, it is worth to review a representative of the data on SMSPDs available in the literature. Table 3 lists the superconductors used to fabricate these devices and compares their geometrical characteristics, operation temperatures, and efficiency (when not available, the presence/absence of the saturation in the PCR is reported) at a specified wavelength. Different materials were tested, from nitrides to amorphous superconductors, all of them intensively investigated to realize high-performing SNSPDs [3, 45]. However, to date, a saturated single photon response in microstrips at 1.55 μm above 1 K was reported only for NbN [18], while WSi [20] and MoSi [19] required to work below this temperature. Moreover, single photon regime was demonstrated at the same wavelength only for NbN [18], and WSi [20]. Notably, He-irradiated NbN devices, shunted with a proper resistance, demonstrated saturated internal detection efficiency at λ=2 μm [22]. As shown in Table 3, a meander geometry is commonly used to guarantee a large effective detection area, but at the price of increasing the values of $L_K$.

Here, we demonstrated single photon sensitivity at λ=1.5 and 2 μm in pairs of short parallel microstrips made of NbRe films at about 1.6 K. We believe that our findings are, on one hand, a consequence of the properties of the electronic system, and, in particular, its higher capability of retaining the photon energy compared to the phonon system, which makes NbRe competitive with amorphous superconductors in the detection at high wavelengths. This occurrence is measured by the parameter γ, as extensively explained in [10]. On the other hand, the value of the superconducting order parameter in NbRe, which is intermediate between NbN and amorphous superconductors, may, in principle assure an operation temperature easily accessible by cryogen-free technology. Moreover, this work is an important milestone towards the realization of large area detectors with low $L_K$, since the device design can be extended from a single pair to a series of parallel microstrips, or to even more complex geometries.

Finally, our efforts are now towards a saturated photo response regime and the extension of the detection wavelength. Both these targets can be in principle achieved by reducing the film thickness and the operation temperature.

**Conclusions**

In conclusion, this work demonstrates the suitability of NbRe for the realization of SMSPDs working beyond 1.55 μm at a relatively high temperature. Single-photon regime operation was clearly demonstrated up to 2 μm on a pair of parallel strips, a geometry never explored before. This innovative layout can be considered the building block of more complex designs which may result in large detection area and low kinetic inductance. Future experiments aim at optimizing the detectors for specific applications and to extend the detection capability even further into longer wavelengths.

**Methods**

**Device patterning**

The SMSPDs were optically patterned by direct writing using a smart print (Microlight3D) starting from the as-deposited NbRe films. The AZ1505 positive resist has been spanned at 4000 rpm for 60 s on the films, to obtain a resist layer with a nominal thickness of 600 nm. This guarantees the realization of strips of reduced widths with a high degree of resolution and reproducibility. Subsequently, the samples have been illuminated by blue (430 nm) light for an exposure time of 0.3 s and then developed for 30 s in AZ351B, which has been diluted in distilled water according to a 1:4 ratio. Great care has been paid to the films exposure due to the reduced working area of the microprinter. In fact, the system can exposure areas of 1920 pixels × 1080 pixels, where one pixel corresponds to 704 nm. Therefore, the realization of large patterns requires the consecutive exposure of the different areas in which the main geometry can be divided. During this process, the movement of the projector from one point of the surface to the other is automatically implemented by a micromanipulator, allowing a shift along the horizontal direction, $x$, and the vertical direction, $y$, by following a path that includes all the dowels. In order to overcome any problems in the connection between the different parts of the drawing, the stitching technique (i.e., the partial overlapping of the adjacent pieces) is used. In the case of series of pairs of parallel microstrips the stitching used introduces a proximity effect with regions which are more exposed with respect to others.

Finally, a purely physical process based on Ar ion etching has been used to remove the NbRe film not exposed. The process is realized at an Ar ion pressure of $5.1 \times 10^{-4}$ kPa, at a power of 4.5 W, and with a cathode current of about 2.2 A. With these parameters the etching rate is of 1nm/min. The poor selectivity of the process is overcome by the large thickness of the resist layer, which is much thicker than the one of the material to be removed. By using this technique, strips with well-defined profiles can be obtained.

**Data availability**

The datasets generated during and/or analysed during the current study are available from the corresponding author on reasonable request.

**Acknowledgments**

The authors wish to thank M. Casalino for the availability of the laser source at 2.0 μm.
This research was supported by the QUANCOM Project 225521 (MUR PON Ricerca e Innovazione No. 2014–2020 ARS01_00734) and by the Project "Quantum Italy Deployment (QUID)" n°101091408 in the framework of the Digital Europe Programme.

| Device | $w$ (μm) | $L$ (μm) | $A$ (μm²) | $I_c$-$I_r$ (μA) | $J_c$ (MA/cm²) |
|---|---|---|---|---|---|
| D1 | 1.4 | 5 | 14 | 6.2 | 0.30 |
| D2 | 2.2 | 7.5 | 33 | 19.2 | 0.14 |
| D3 | 2.1 | 10 | 42 | 8.2 | 0.27 |

Table 1. Fabricated devices based on a pair of parallel strips, their geometrical characteristics, along with their transport and superconducting properties. $I_c$ and $I_r$, are the switching and the retrapping current, respectively, as shown in Fig. 2. All values refer to T=4.2 K.

| λ (μm) | $I_b/I_c$ | $b$ |
|---|---|---|
| 1.55 | 0.85 | $1.11 \pm 0.02$ |
| 1.55 | 0.90 | $1.13 \pm 0.06$ |
| 1.55 | 0.95 | $0.93 \pm 0.09$ |
| 2.0 | 0.90 | $0.99 \pm 0.01$ |

**Table 2.** Values of the exponent $b$ extracted from the fitting procedure of the count rate as a function of the normalized light intensity for different current bias and wavelengths.

| Material | Geometry | width (μm) | thickness (nm) | operation temperature (K) | wavelength (μm) | Efficiency/ count rate | single-photon regime |
|---|---|---|---|---|---|---|---|
| NbN [18] | Constriction-type bridge | 2.12 | 5.8 | 1.7 | 1.55 | 10% | yes |
| NbN [21] | meander | 1 | 7 | 0.84 (2.1) | 1.55 | 92.2% (70%) | not available |
| He irradiated NbN [22] | double-spiral | 1 | 7 | 0.85 | 1.55 (2.0) | saturated (saturated) | not available |
| NbTiN [44] | meander | 1 | 6 | 2.2 | 1.55 (0.85) | not saturated (saturated) | not available |
| MoSi [19] | meander | 1-3 | 3-5 | 0.3 | 1.55 | not available (saturated) | not available |
| MoSi [46] | meander | 1 | 2 | 0.8 | 1.065 | not available (saturated) | not available |
| WSi [20] | meander | 1-3 | 2.1-2.8 | 0.8 | 1.55 | not available (saturated) | yes |
| WSi [46] | meander | 1 | 2.2-5.1 | 0.8 | 1.065 | not available (saturated) | not available |

**Table 3.** Overview of the main characteristics of a representative of SMSPDs based on different superconductors.